# Independent Perceptual Process of Microscopic Texture and Surface Shapes through Lateral Resistive Force Cues

Mirai Azechi[1] and Shogo Okamoto[1]

[1] *Department of Computer Science. Tokyo Metropolitan University, Hino, Japan*

Email: okamotos@tmu.ac.jp

**Abstract**—Macroscopic surface shapes, such as bumps and dents, as well as microscopic surface features, like texture, can be identified solely through lateral resistive force cues when a stylus moves across them. This perceptual phenomenon has been utilized to advance tactile presentation techniques for surface tactile displays. However, the effects on shape recognition when microscopic textures and macroscopic shapes coexist have not been thoroughly investigated. This study reveals that macroscopic surface shapes can be recognized independently of the presence of microscopic textures. These findings enhance our understanding of human perceptual properties and contribute to the development of tactile displays.

**Keywords:** Surface shape, texture, bump, dent

## 1 Introduction

Surface shapes such as bumps and dents can be discriminated through lateral resistive force cues without requiring vertical finger displacement or geometrical cues [1,2]. By utilizing this perceptual phenomenon, it is possible to present three-dimensional surface shapes on a flat surface [3–7], and the perceptual characteristics of these virtual surface shapes have been investigated [8]. However, when macroscopic surface shapes and microscopic textures coexist, how they interfere with each other is unclear.

Our previous research examined the recognition accuracy of macroscopic bumps with microscopic textures combined on an electrostatic friction display, demonstrating that each shape can be identified when they coexist [9]. However, the effects of microscopic textures on the recognition of macroscopic bumps have not been thoroughly investigated.

This study aims to investigate how microscopic textures affect the recognition of macroscopic surface shape when they coexist. There are three hypotheses regarding the potential influence of microscopic textures on the perceptual properties of macroscopic bumps.

The first hypothesis is that the presence of texture deters the recognition of surface shapes. Specifically, the stimuli produced by microscopic textures might be more prominent than those from macroscopic bumps, possibly interfering with the recognition of macroscopic surface shapes.

The second hypothesis is that the presence of texture does not deter the recognition of surface shapes. It might be possible to appropriately separate the low-frequency components produced by macroscopic bumps from the high-frequency components generated by microscopic textures, allowing for the recognition of surface shapes. Our previous study [9] supports this hypothesis, as it demonstrated the ability to distinguish between textures and bumps.

The third hypothesis is that the presence of the texture deters the recognition of dents and fosters the perception of bumps. In this experiment, the texture is placed only near the center of the surface shape. As illustrated in Section 2.2, the resistive force initially increases for bump and decreases for dent. Therefore, the temporary rise in resistive force at the transition from a smooth surface to texture may enhance or bias the recognition of bumps and hinder the recognition of dents.

Until now, the influence of such textures on surface shape recognition has not been investigated. Therefore, by testing these hypotheses, we aim to deepen our understanding of the perception of both microscopic and macroscopic surface features. Surface recognition based on lateral force cues underly the friction-variable surface

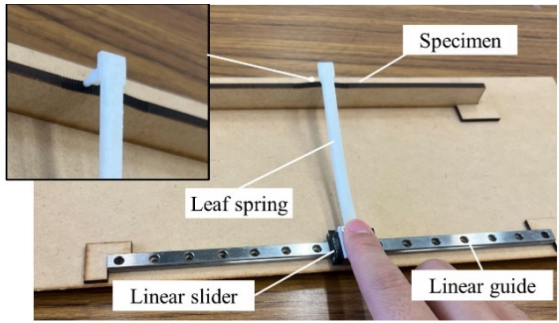

Fig. 1 Apparatus used to decouple the vertical displacement and lateral force. A plastic sphere at the tip of the leaf spring is in constant with the surface.

texture displays [10–13], and the findings of this study will contribute to their advancement.

## 2 METHODS

### 2.1 Apparatus: Deliver lateral resistance force

In our experiments, we employed a lateral force delivering system, as shown in Fig. 1. The apparatus consisted of a leaf spring attached to a linear slider, which moved horizontally. A plastic sphere at the end of the leaf spring maintained continuous contact with a surface specimen. As participants operated the linear slider, only lateral force was conveyed to their fingers, ensuring no vertical movement. This design allowed us to replicate friction-variable tactile displays with high precision. A comparable apparatus was used in our previous research [14].

### 2.2 Stimuli: Bump and dent with or without texture

As shown in Fig. 2, we used a specimen with a Gaussian bump or dent, with or without texture, and controlled the maximum gradient and width across two levels, outlined in Table 1. Thus, eight specimens were used in total (2 gradient levels × 2 width levels × 2 shapes).

In the absence of texture, the maximum gradient determines the peak lateral force on the finger, while the width influences the rate of force change. A steeper gradient and a narrower width result in a more rapid change in force. For bumps, the lateral force first increases while rising on the slope, then decreases while descending. On the other hand, the force for dents first decreases, then increases. Therefore, the difference in these force changes lies in the order.

As shown in Fig. 3, for bump and dent shapes with texture, a sinusoidal microscopic texture was placed on the center of the specimen, from the beginning of the slope to the end of the slope. The amplitude of the texture was adjusted so that the peaks of the sinusoidal waves coincided with the Gaussian bump and dent surfaces. For those surface with texture, the lateral resistive force variation is larger compared to the surfaces without texture.

| Table 1 Parameters of bump and dent surface. | | |
|---|---|---|
| Maximum gradient (-) | 0.079 | 0.125 |
| Width (mm) | 5.0 | 9.0 |

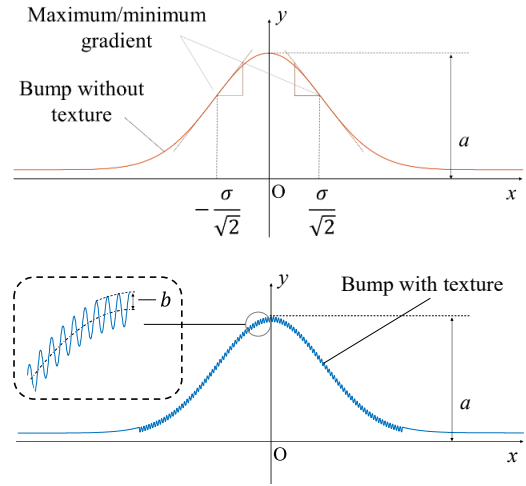

Fig. 2 Parameters of the surface. The top figure shows the bump surface without texture, the bottom figure shows the bump surface with texture. These bumps become dents when inverted along the x-axis.

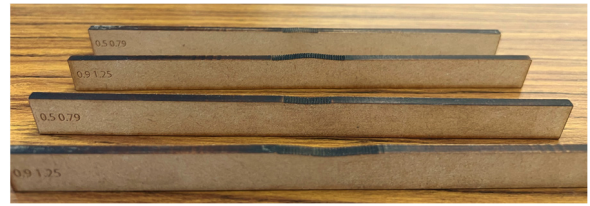

Fig. 3 Wooden specimen with a bump or dent with texture at the center.

The Gaussian function and sinusoidal texture of the surfaces was defined as:

$$y(x) = \pm a\left(\exp\left(-\frac{x^2}{\sigma^2}\right)\right) + b\sin(x) - b \quad (1)$$

where $a$, and $\sigma$ denote the height and the width of the surfaces, respectively. The amplitude of microscopic texture, represented as $b$, was 0.2 mm. The $b$ value was 0 when the texture was not present.

Fig. 3 shows the actual bump and dent shapes with texture, cut from medium-density fiberboard using a laser cutter (VLS 6.60, Universal Laser Systems, Inc., USA). The cut surfaces were refined using fine abrasive papers.

Table 2  Mean proportions and standard errors of correct answer for each surface.

| Surface shape | Width (mm) | Maximum gradient (-) | Mean correct proportions ± standard errors |
|---|---|---|---|
| Bump | 5.0 | 0.079 | 0.62 ± 0.06 |
| Bump + texture | 5.0 | 0.079 | 0.69 ± 0.08 |
| Bump | 9.0 | 0.125 | 0.80 ± 0.04 |
| Bump + texture | 9.0 | 0.125 | 0.89 ± 0.04 |
| Dent | 5.0 | 0.079 | 0.70 ± 0.04 |
| Dent + texture | 5.0 | 0.079 | 0.64 ± 0.07 |
| Dent | 9.0 | 0.125 | 0.78 ± 0.05 |
| Dent + texture | 9.0 | 0.125 | 0.79 ± 0.08 |

### 2.3 Experimental procedures

During the training session, participants examined visible samples of bumps and dents distinct from those used in the main experiment. This allowed them to determine optimal hand speeds for shape recognition. Each participant spent approximately five minutes in this initial phase.

In the main session, participants, blindfolded and wearing earmuffs to eliminate auditory cues, performed a two-alternative forced-choice task. They responded to eight different surface shapes presented randomly. Each shape was presented 10 times, totaling 80 trials. For each trial, participants could explore the surface twice—once from left to right and once from right to left. After the exploration, participant answered whether the surface was a bump or dent in a two-alternative forced choice manner.

### 2.4 Participants

Seven volunteers participated in this experiment without prior knowledge of the research objectives.

### 2.5 Data analysis

To examine the difference in the correct answer proportions between each surface shape with and without textures, we performed a Wilcoxon signed-rank test on the mean correct answer proportions of each surface shapes across all participants.

### 3 RESULTS

Table 2 lists the means and standard errors of the proportions of correct responses for each surface shape across all participants. For the bump with a width of 9.0 mm and a maximum gradient of 0.125, a Wilcoxon signed-rank test revealed a significant difference in the correct response proportions between the conditions with and without texture ($T = 0$, $p < 0.001$). The shape was recognized with higher accuracy when the texture was present.

No significant differences were observed for the other bumps and dents: bump with a width of 5.0 mm and a maximum gradient of 0.079 ($T = 5$, $p = 0.14$); dent with a width of 5.0 mm and a maximum gradient of 0.079 ($T = 8.5$, $p = 0.14$); and dent with a width of 9.0 mm and a maximum gradient of 0.125 ($T = 11$, $p = 0.32$).

### 4 DISCUSSION

The experimental results indicate that microscopic texture does not impede the recognition of macroscopic bumps and dents. It is suggested that the low-frequency components produced by macroscopic shapes can be recognized separately from the high-frequency components generated by microscopic textures, allowing for shape identification.

On the other hand, it is noted that for bump with the width of 9.0 mm and maximum gradient of 0.125, the textured surface was more correctly recognized compared to the one without texture. One possible explanation is that the temporary rise in resistive force, during the transition from a smooth surface to texture, was interpreted as the rising slope of a bump, enhancing the perception of bump shape. However, there are some doubts about this explanation. If the temporary increase in resistive force fosters bump recognition, the correct answer proportions for the other bump shape should similarly increase because of the texture as well. Moreover, using the same reasoning, the correct answer proportions for dent shapes would decrease. However, no such consistent trend was observed in these results. The impact of texture placement on the perception of surface shapes needs to be investigated in more detail under more specific conditions in future studies.

### 5 CONCLUSION

This study investigated the impact of microscopic textures on the recognition of surface shapes when only lateral force cues were available. Our results indicate that microscopic textures neither hinder nor enhance the recognition of macroscopic surface shapes, such as bumps and dents. Although the texture covered the entire shape in this study, the effects of its local placement remain unknown. Future research in this area will contribute to a deeper understanding of human perceptual properties and advance haptic presentation techniques for surface tactile displays.

ACKNOWLEDGEMENT

This work was supported in part by MEXT Kakenhi (24K03019 and 23H04360).